\documentclass[pra,aps,twocolumn]{revtex4}

\usepackage{color} 
\usepackage{graphicx} 

\begin{document}
\title{Competition  of superfluidity and density waves 
in one-dimensional Bose-Fermi
  mixtures}
\author{E. Orignac}
\affiliation{Laboratoire de Physique de l'ENS-Lyon, CNRS UMR 5672, 46
  All\'ee d'Italie, 69364 Lyon Cedex 07, France} 
\author{M. Tsuchiizu} 
\author{Y. Suzumura} 
\affiliation{Department of Physics,  Nagoya University, 
 Nagoya 464-8602, Japan}
\begin{abstract}
We study a mixture of one-dimensional 
bosons and spinless fermions at incommensurate
filling using phenomenological bosonization and Green's functions
techniques. We derive the relation between the parameters of the
microscopic Hamiltonian and macroscopic observables. Galilean
invariance results in extra constraints for the current current
interactions. 
We obtain  the exact exponents for the various response functions, 
and show that
superfluid fluctuations are enhanced by the effective boson-fermion 
density-density  interaction and suppressed by the effective boson-fermion
current-current interaction. In the case of a bosonized model with purely
density-density interaction, when the effective  boson-fermion
density-density interaction  is
weak enough, the superfluid exponent of the bosons 
has a non-monotonous variation 
with the ratio of the fermion velocity to the boson velocity.  
By contrast, density-wave exponent 
and the exponent for fermionic
superfluidity are monotonous functions of the velocity ratio.    
\end{abstract}
\date{\today}
\maketitle

\section{Introduction} 

The recent development of atom trapping
technologies \cite{grimm_potential_review,reichel} has led to the opening 
of new research fields of many-body physics. In particular, the
possibility of controlling to a large degree 
the shape of the trapping potential has permitted the experimental
realization of nearly isolated low dimensional interacting boson
systems \cite{koehl_1dbose}, and the observation of the bosonic Mott
insulator \cite{koehl_mott_review}. Moreover, the use of Fano-Feshbach
resonances \cite{feshbach_resonance,fano_resonance} allows
experimentalist  to control interaction strength in the 
low dimensional atomic gases
\cite{courteille98_feshbach_rb,roberts98_feshbach_rb,inouye98_feshbach_na}.
This has permitted the observation of the
Girardeau \cite{girardeau_bosons1d} fermionization of one-dimensional
bosons with strongly repulsive
interactions \cite{paredes_toks_experiment,kinoshita_tonks_experiment}. 
From the theoretical point of view, fermions as well as 
bosons with repulsive interactions
in one dimension are in the Tomonaga-Luttinger liquid
state \cite{haldane_bosons,giamarchi_book_1d,cazalilla_correlations_1d,petrov04_bec_review}. The
bosonization technique \cite{haldane_bosons} allows the description of
their low-energy physics in terms of collective phonon-like modes as
well as the computation of the various correlation functions. It is
found that in the Tomonaga-Luttinger liquid state at zero temperature,
only quasi-long range order (whether superfluid or density wave) can
be obtained, with correlation functions decaying as power laws, the
exponent of the power law being a rational function of the Tomonaga-Luttinger
parameter $K$. This parameter $K$ depends on interaction. In  the case
of bosons with contact interaction \cite{lieb_bosons_1D}, this
parameter can be calculated exactly, and it is found that $K\ge 1$
making superfluidity the dominant correlation. When the interaction
becomes extremely repulsive, or the density becomes very low, the
Girardeau model is recovered \cite{girardeau_bosons1d} and the exponent
$K$ goes to 1. In the case of spinless fermions, perturbative
calculations \cite{dzyaloshinskii_larkin,solyom_RG_multiplicatif,luther_bosonisation}
show that the exponent $K$ is smaller than 1 for repulsive interactions
(thus favoring density waves) and larger than 1 for attractive
interactions (thus favoring a $p$-wave type superfluidity). 
For finite temperatures,
the correlation functions decay exponentially, with a correlation
length inversely proportional to temperature. Such behavior gives rise
to susceptibilities that diverge as a power law with temperature, with
an exponent determined by the zero temperature correlation exponent.  
Recent
experiments \cite{schreck_mixtures_optical,inouye_fbat,guenter2006,mcnamara2006,best2009} with trapped atoms have shown the possibility of realizing
many-body systems mixing fermions and bosons. From the theoretical
point of view, mixtures of bosons and spinless fermions in one
dimension were considered
in Refs.\ \onlinecite{cazalilla03_mixture,mathey04_mix_polarons,mathey2007a}. It was
found that sufficiently strong repulsion or attraction between
fermionic and bosonic atoms could result in respectively a phase
separation or a collapse, while interactions of moderate strength 
would lead (in the case of incommensuration between the atom
densities) to the formation of a two-component Tomonaga-Luttinger
liquid. In Refs.\ \onlinecite{mathey04_mix_polarons,mathey2007a}, it was proposed
to describe the two-component Tomonaga-Luttinger liquid in terms of 
polarons, and various exponents of correlation functions were
obtained. An integrable model of interacting bosons and fermions was
considered in
Refs.\ \onlinecite{imambekov2006a,imambekov2006,frahm05_bosefermi1d,batchelor2005,yin2009}, 
in which a  two-component Tomonaga-Luttinger liquid was found. 
The mixture has also been investigated in numerical
calculations \cite{takeuchi2005a,takeuchi2005,takeuchi2007,rizzi2008,pollet2006,pollet2008,zujev2008}. 
More recent experimental work has considered mixtures of fermionic
atoms. Motivated by these experiments, 
there has been some numerical studies of mixtures of three fermionic species
\cite{luscher2009}.

Although the one-dimensional
 boson fermion mixture has been studied extensively including 
  analytical expression of the
  exponents\cite{mathey04_mix_polarons,mathey2007a}  
of the equal time correlation function, the previous work has
emphasized the polaron correlations over the density wave
correlations. It is thus worthwhile to analyze the competition 
between density-wave and superfluid correlation as interactions in the
system are varied. Besides the question of the effect of interaction
on the exponents, an important question relevant for experiment is the
evolution of the exponents as a function of relative concentration of
fermions and bosons in the mixture. Fermions could have two opposite
effects. On the one hand, fermionic particles can give rise to
retarded attractive interactions between the bosons, which can
compensate the direct boson-boson repulsion and favor superfluidity
and possibly collapse of the mixture. On the other hand, fermionic
particles do not form Bose condensates, and may thus disfavor the
formation of a superfluid state. {For} the  simple Bose-Fermi mixture
model, it is useful to examine the effect of the doping  
on the correlated states of respective particles, e.g. ,  
how the superfluidity is enhanced or suppressed by the interaction
between boson and fermion. A last question, in order to compare
numerical and analytical result, is to develop a phenomenological
bosonization scheme allowing for the calculation of the parameters of
the low energy effective Hamiltonian from ground state energy
calculations.

In the present article, we present a study of mixtures of fermionic or
bosonic atoms using bosonization and a Green's function
formalism. After discussing briefly in Sec.~\ref{sec:hamiltonian} the
phenomenological bosonized 
Hamiltonian of the model and its relation with the macroscopic observables, we
focus in Sec.~\ref{sec:correlations} on the calculation by Green's
function techniques of the 
exponents for superfluidity and density waves. We find a relation 
between the products of density wave and superfluid exponents for 
the bosons and the fermions. We also give an expression of the
correlation lengths at finite temperature. 
In Sec.~\ref{sec:bragg-scattering}, we use the Green's function
technique to obtain the Bragg scattering intensity
following Ref.~\onlinecite{brunello_bragg}. Then, in Sec.~\ref{sec:toy-model}, 
we consider the effect of the variation of the density of one of the
two atom species first in the case of the model of
Ref.~\onlinecite{cazalilla03_mixture}. We show that for this model, the
variation of the bosonic superfluid exponent with fermion density is
not monotonous in the case of a weak interaction. { We also consider the
effect of the effective current-current interaction and show that it
suppresses superfluidity.}          

\section{Hamiltonian}\label{sec:hamiltonian}
We consider a mixture of spinless fermions and bosons. The Hamiltonian
reads: 
\begin{eqnarray}
  H&=&\int dx \left[ -\frac{\hbar^2}{2M_B} \psi^\dagger_B \partial_x^2
    \psi_B  -\frac{\hbar^2}{2M_F} \psi^\dagger_F \partial_x^2 \psi_F
  \right] \nonumber \\ 
&+& \frac 1 2 \int dx dx' [V_{FF}(x-x') \rho_F(x) \rho_F(x')
\nonumber
  \\ &+&
  V_{BB}(x-x') \rho_B(x) \rho_B(x') +2 V_{BF}(x-x') \rho_B(x)
  \rho_F(x')],  
  \label{eq:hamiltonian}
\end{eqnarray}
where $M_F,M_B$ are the masses of (respectively) the fermionic and
bosonic atoms, $V_{FF},V_{BB},V_{BF}$ (respectively) the
fermion-fermion, boson-boson and fermion-boson interactions,
and $\psi_F,\psi_B$ the (respectively) bosonic and fermionic
annihilation operators.  We have
also defined the density operators: $\rho_\nu=\psi^\dagger_\nu
\psi_\nu$. The model~(\ref{eq:hamiltonian}) can be treated by the
method of bosonization \cite{giamarchi_book_1d}.

When
the boson and fermion densities are not commensurate with each other,
the bosonized Hamiltonian reads~\cite{cazalilla03_mixture,mathey04_mix_polarons,mathey2007a}:  
\begin{eqnarray}
  \label{eq:bosonized}
  H=\sum_{a,b} \int \frac{dx}{2\pi} \left[ \pi^2 M_{ab} \Pi_a \Pi_b +
    N_{ab} \partial_x \phi_a \partial_x \phi_b \right],   
\end{eqnarray}
where $[\phi_a(x),\Pi_b(x')]=i \delta_{a,b} \delta(x-x')$ and $a,b \in
\{B,F\}$. Indeed, the Hamiltonian (\ref{eq:hamiltonian}) is invariant
under a parity transformation. Under such a transformation,  
$\phi_a(x) \to -\phi_a(-x)$ and $\Pi_a(x) \to -\Pi_a(-x)$, 
so that quadratic terms $\Pi_a \partial_x \phi_b$ change sign under
parity, and Eq.~(\ref{eq:bosonized}) is the most general Hamiltonian
quadratic in $\Pi_a$ and $\partial_x \phi_a$ for $a=B,F$. 
The matrices $M_{ab}$ and $N_{ab}$ in Eq.\ (\ref{eq:bosonized}) are
real symmetric and are deduced from variations 
of the ground state energy $E_{GS}$ of a finite system of size $L$  
with (respectively)  changes of boundary
conditions $\psi_a(L)=e^{i\varphi_a} \psi_a(0)$  and changes of
particle densities $\rho_a=N_a/L$:       
\begin{eqnarray}
 \label{eq:definition-M}
    M_{ab} = \pi L \frac{\partial^2
    E_{GS}}{\partial\varphi_a \partial\varphi_b}, \\
  \label{eq:definition-N}
  N_{ab} = \frac 1 {\pi L}  \frac{\partial^2
    E_{GS}}{\partial\rho_a \partial\rho_b}.  
\end{eqnarray} 
In the case of a Galilean invariant Hamiltonian such
as Eq.\ (\ref{eq:hamiltonian}),  it is possible to
further constrain the elements of the matrix $M$ with the relations
(see App.~\ref{app:galilean}): 
\begin{eqnarray}\label{eq:galilean-constraint} 
\pi \rho_F &=& M_{FF} M_F + M_{BF} M_B, \\
   \pi \rho_B &=& M_{BF} M_F + M_{BB} M_B.
\end{eqnarray}
So that knowing just one of the
parameters $M_{FF}$, $M_{BB}$, and $M_{BF}$  fully determines the matrix
$M$. If one starts from the limit $V_{BF}=0$ and applies bosonization
first, and then reintroduces the interaction $V_{BF}(x-x') =V_{BF}
\delta(x-x')$ as a perturbation, to lowest order one obtains the
Hamiltonian:
\begin{eqnarray}
  \label{eq:bosonized-perturb}
  H&=&\sum_{\nu=F,B} \int \frac{dx}{2\pi} \left[ u_\nu K_\nu(\pi \Pi_\nu)^2 +
    \frac{u_\nu}{K_\nu} (\partial_x
    \phi_\nu)^2\right] \nonumber \\ 
&+& \frac{V_{BF}}{\pi^2} \int dx  \partial_x
  \phi_B \partial_x \phi_F,  
\end{eqnarray}
in which  $u_\nu K_\nu = \frac{\pi \rho^{(0)}_\nu}{M_\nu}$ as a result
  of Galilean invariance of Eq.\ (\ref{eq:hamiltonian}), 
$\frac{u_\nu}{K_\nu}=\frac{1}{\pi (\rho^{(0)}_\nu)^2 \kappa_\nu}$ 
where $\rho^{(0)}_\nu$ is the
density of particles, and $\kappa_\nu$ is the compressibility.
 In the case of non-interacting
fermions $V_{FF}=0$, one has $K_F=1$. 
For hard core bosons \cite{girardeau_bosons1d},  $K_B=1$, while 
for the Lieb-Liniger model \cite{lieb_bosons_1D}, $V_{BB}(x-x')=V_{BB}
\delta(x-x')$,   
one has $K_B\ge 1$, with $K_B
\to \infty$ when $V_{BB} \to 0$  and $K_B \to 1$ for $V_{BB} \to
\infty$.  Note that at this low order in perturbation theory, no
term $M_{BF}$ is present. This term is expected to appear in second
order perturbation theory, along with corrections to the bare terms
$M_{FF}$ and $M_{BB}$. A Hamiltonian with quadratic interactions
similar to Eq.\ (\ref{eq:bosonized}) but also comprising interband
tunneling terms was considered in the context of a two band model of
interacting spinless fermions \cite{muttalib1986}.  
 A Hamiltonian equivalent to
Eq.\ (\ref{eq:bosonized-perturb}) has been studied by path integral methods  
as a model of  one-dimensional electrons
interacting with acoustic phonons \cite{loss1994,martin1995}.
Due to the quadratic character of the
Hamiltonian~(\ref{eq:bosonized}), its spectrum is readily
obtained \cite{cazalilla03_mixture} as two branches
$\omega_\pm(q)=u_\pm |q|$ with linear spectrum, 
showing that its ground state
is a two component Tomonaga-Luttinger liquid provided the velocities $u_\pm$ of both 
components are real. The vanishing of the velocity $u_-$ 
is an indication of instability \cite{cazalilla03_mixture}
either towards phase separation (in the case of repulsive
boson-fermion interaction) or collapse (in the case of attractive
boson-fermion interaction). 

\section{Superfluid and density wave correlations}
\label{sec:correlations}

From the diagonalization of  the Hamiltonian~(\ref{eq:bosonized}),
bosonization allows to obtain
the exponents of the various correlation
functions \cite{mathey2007a}. In this paper, we will use a
different (but equivalent) approach to compute the correlation 
functions of the system. 

 By an equation of motion method, we 
first obtain the Green's functions:
\begin{eqnarray}
  \label{eq:Green1}
  G_{ab}(x,\tau)=-\langle T_\tau (\phi_a(x,\tau)-\phi_a(0,0))
  \phi_b(0,0) \rangle, \\ 
 \label{eq:Green2} 
  \bar{G}_{ab}(x,\tau)=-\langle T_\tau (\theta_a(x,\tau)-\theta_a(0,0))
  \theta_b(0,0) \rangle,
\end{eqnarray}
where 
$\theta_a=\pi \int^x dx'  \Pi_a(x')$
 and $a,b$ can be $B$ or
$F$. From these Green's functions
(\ref{eq:Green1}) and (\ref{eq:Green2}), 
we find the correlation functions of exponential
fields as:
\begin{eqnarray}
\label{eq:exponentials} 
  \left \langle T_\tau e^{i\sum_a \lambda_a \phi_a(x,\tau)}  e^{-i\sum_a
    \lambda_a \phi_a(0,0)}\right \rangle &=& e^{-\sum_{a,b} \lambda_a
    \lambda_b G_{ab}(x,\tau)} , \\
\label{eq:exponentials2} 
  \left \langle T_\tau e^{i\sum_a \lambda_a \theta_a(x,\tau)}  e^{-i\sum_a
    \lambda_a \theta _a(0,0)}\right \rangle &=&  e^{-\sum_{a,b} \lambda_a
    \lambda_b \bar{G}_{ab}(x,\tau)},
\end{eqnarray}
where $\lambda_B$ and $\lambda_F$ are real numbers.
Such method is of course applicable to cases with more than two
components, as long as the Hamiltonian remains quadratic in the
fields $\phi_a$ and $\theta_a$.

To derive the Green's functions, we start from 
the equations of motions in Matsubara time of the fields $\phi_a$ and
$\Pi_a$ read: 
\begin{eqnarray}\label{eq:eom-fields} 
  \partial_\tau \phi_a(x,\tau) &=&[H,\phi_a]= -i \pi \sum_b M_{ab}
  \Pi_b(x,\tau), \\
  \partial_\tau \Pi_a(x,\tau)&=&[H,\Pi_a]= -\frac i \pi \sum_b N_{ab}
  \partial^2_x \phi_b. 
\end{eqnarray}
The equations of motions for the Green's functions $G_{ab}(x,\tau)$
thus read: 
\begin{eqnarray}
  \label{eq:eom-green}
  \partial_\tau G_{ab}(x,\tau)&=&i \pi \sum_c M_{ac} \langle T_\tau
  \Pi_c(x,\tau) \phi_b(0,0)\rangle, \\ 
  \partial^2_\tau    G_{ab}(x,\tau)&=& \pi \delta(x)\delta(\tau)
  M_{ab} - \sum_c (M N)_{ac} \partial_x^2 G_{cb}(x,\tau). \nonumber
  \\  
\end{eqnarray}
Going to  Fourier space, we obtain ($a,b=B ,F$): 
\begin{eqnarray}
  \label{eq:green-phi-Fourier}
  G_{ab}(q,\omega_n)&=&-\pi ((\omega_n^2  + (M N) q^2)^{-1} M)_{ab} ,
  \end{eqnarray}
where $\omega_n = 2n \pi T$.
From  Eq.\ (\ref{eq:green-phi-Fourier}):
\begin{eqnarray}\label{eq:fourier-def} 
  G_{ab}(x,\tau)=\frac 1 \beta \sum_{\omega_n} \int \frac{dq}{2\pi}
  G_{ab}(q,\omega_n)(e^{iqx -i\omega_n \tau} -1)
  e^{-|q|\alpha},\nonumber \\     
\end{eqnarray}
where we have introduced the cutoff $\alpha$. 
By using the duality transformation $\partial_x \phi_a = \pi P_a$,
$\partial_x \theta_a = \pi \Pi_a$, we obtain  equations of
motion for $\bar{G}_{ab}$ similar to Eq.\ (\ref{eq:eom-green}) with the
roles of $M$ and $N$ exchanged. 
Thus, in Fourier space, we have:
 \begin{eqnarray}
  \label{eq:green-theta-Fourier}
  \bar{G}_{ab}(q,\omega_n)&=&-\pi ((\omega_n^2  + (N M) q^2)^{-1}
  N)_{ab}. 
\end{eqnarray}
The expressions (\ref{eq:green-phi-Fourier}) and
(\ref{eq:green-theta-Fourier}) show that the retarded Green's
functions have  poles for $i \omega$ equal to $u_\pm |q|$ 
where $u_-^2 \le u_+^2$ are the two eigenvalues of $MN$. Stability
requires that $u_\pm^2 >0$ and thus $\mathrm{det}(MN)>0$. Since we
know from App.~\ref{app:galilean} that $\det(M)>0$ in a Galilean
invariant model, instabilities occur for $\mathrm{det}(N)=0$. From the
definition of $N$, Eq.~(\ref{eq:definition-N}), such instabilities are
either collapse or phase separation.        

Further, for zero temperature and $\tau=0$, we can obtain a general  form 
for
$G$ using Eq.\ (\ref{eq:fourier-def}). We find: 
\begin{eqnarray}
  G(x,0)&=&\frac 1 2 (M N)^{-1/2} M \ln \left(\frac{\sqrt{x^2+\alpha^2}} \alpha \right), \\  
  \bar{G}(x,0)&=&\frac 1 2 (NM)^{-1/2} N \ln
  \left(\frac{\sqrt{x^2+\alpha^2}} \alpha \right). 
\end{eqnarray}
Then equations (\ref{eq:exponentials}) and (\ref{eq:exponentials2}) lead to:
\begin{eqnarray}
  \langle T_\tau e^{i \sum_a \lambda_a \phi_a(x,0)} e^{-i \sum_a
    \lambda_a \phi_a(0,0)} \rangle =
\left(\frac{\alpha}{\sqrt{x^2+\alpha^2}}\right)^{\frac 1 2 {}^t\lambda (MN)^{-1/2} M
  \lambda}, \\ 
 \langle T_\tau e^{i \sum_a \lambda_a \theta_a(x,0)} e^{-i \sum_a
    \lambda_a \theta_a(0,0)}\rangle =
\left(\frac{\alpha}{\sqrt{x^2+\alpha^2}}\right)^{\frac 1 2 {}^t\lambda (NM)^{-1/2} N
  \lambda},
\end{eqnarray}
where ${}^t\lambda=(\lambda_B,\lambda_F)$.
We can define the matrices $\eta_\phi = (MN)^{-1/2} M$ and
$\eta_\theta= (NM)^{-1/2} N=(NM)^{1/2} M^{-1}$. 
These matrices yield the exponents for
the exponential fields. One can see that the duality relations become
$\eta_\phi {}^t\eta_\theta=\eta_\theta {}^t\eta_\phi = 1$. For the two
component system, one has the identity: 
\begin{eqnarray}
  \label{eq:matrix-square-root}
  (MN)^{-1/2}=\frac{I}{u_++u_-} + \frac{u_+ u_-}{u_++u_-} (MN)^{-1}, 
\end{eqnarray}
which can be checked by applying the right hand side of the formula 
to each eigenvector of $MN$. We thus have:
\begin{eqnarray}
  \eta_\phi&=&\frac{1}{u_++u_-} (M + u_+ u_- N^{-1}) \\
  \eta_\theta&=&\frac{1}{u_++u_-}(N+u_+ u_- M^{-1})  
\end{eqnarray}
In the case where $\det(N) \to 0$, we see that $\eta_\phi$ will have
matrix elements going to infinity as $u_-^{-1}$, 
whereas $\eta_\theta \to N/u_+$. 
Therefore, near a collapse or a phase separation the density wave
exponents are divergent, while the superfluid exponents remain
finite. 

For nonzero temperature, the sum Eq.~(\ref{eq:fourier-def}) is dominated
for long distances by the term with $n=0$. One finds that:
\begin{eqnarray}
  \label{eq:thermal-green}
  \frac{1}{\beta} \int G(q,\omega_0=0) (e^{iqx}-1) \frac{dq}{2\pi} &=&
  \frac \pi 2 T |x| N^{-1}, \\
 \frac{1}{\beta} \int \bar{G}(q,\omega_0=0) (e^{iqx}-1) \frac{dq}{2\pi} &=&
  \frac \pi 2 T |x| M^{-1}, 
\end{eqnarray}
so that the correlation functions decay exponentially for long
distances, 
\begin{eqnarray}
  \langle T_\tau e^{i \sum_a \lambda_a \phi_a(x,0)} e^{-i \sum_a
    \lambda_a \phi_a(0,0)} \rangle \sim 
e^{-\frac {\pi T} 2  {}^t\lambda N^{-1} 
  \lambda |x|}, \\ 
 \langle T_\tau e^{i \sum_a \lambda_a \theta_a(x,0)} e^{-i \sum_a
    \lambda_a \theta_a(0,0)}\rangle =e^{-\frac {\pi T} 2  {}^t\lambda M^{-1} 
  \lambda |x|},   
\end{eqnarray}
with thermal correlation lengths given respectively by
$\xi_\phi(\lambda)=2/(\pi T  {}^t\lambda (N)^{-1} 
  \lambda)$ and $\xi_\theta(\lambda)=2/(\pi T  {}^t\lambda (M)^{-1} 
  \lambda)$. We note that near the instability, the correlation length
  $ \xi_\phi(\lambda)$ goes to zero, in accordance with the reduction
  of density wave ordering found at zero temperature, while the length
  $\xi_\theta$ remains finite. 
\color{black}

\subsection{Atomic density wave correlations}
\label{sec:density}

To characterize the atomic density-wave (ADW)  ordering of the
fermions, 
 the field operator of which is given by:  
\begin{equation}\label{eq:fermion-bosonized} 
\psi_F(x)=e^{i k_F x} \psi_F^+ + e^{-i k_F x} \psi_F^-, \quad
\psi_F^{\pm} \sim \frac{{\rm e}^{i( \pm \phi_F +
    \theta_F)}}{\sqrt{2\pi\alpha}}, 
\end{equation}
where $k_F=\pi \rho_0^{(F)}$ is the Fermi wavevector, $\alpha$ is a
short-distance cutoff,  and $\pm$ label
the two Fermi points, we have to
calculate the correlation function of the $2k_F$ component of the
density operator $\psi^\dagger(x) \psi(x)$, namely  
$(\rho_{2k_F} \sim (\psi_F^{+})^{\dagger} \psi_F^{-})$:
\begin{eqnarray}\label{eq:fermion-density-correlations} 
  \langle \rho_{2k_F}^\dagger(x) \rho_{2k_F}(0)\rangle 
 \sim \langle {\rm e}^{- i 2 \phi_F(x) + i 2 \phi_F(0)} \rangle 
\sim (x/\alpha)^{-\eta^{(F)}_{ADW}}.  
\end{eqnarray}
The density wave exponent $\eta^{(F)}_{ADW}$ is: 
\begin{eqnarray}
  \label{eq:exponent-adw-f}
  \eta^{(F)}_{ADW}=\frac{2}{u_++u_-}\left[M_{FF} + \frac{u_+
    u_-}{\mathrm{det}(N)} N_{BB}\right] .
\end{eqnarray}
For $V_{BF}=0$, the exponent of Eq.\ (\ref{eq:exponent-adw-f}) reduces to
$2K_F$. Near the collapse (for $V_{BF}<0$) or the phase
separation (for $V_{BF}>0$) which is obtained at $u_{-} \rightarrow 0$, 
we note that the exponent
$\eta_{ADW}^{(F)}$ is diverging as $\sim 1/u_-$.   
The fermionic density-density correlation at small wavevectors can
also be obtained from bosonization. We have:
\begin{eqnarray}
  \label{eq:fermion-density-corr-uf}
  \langle \rho_{F,0}(x)  \rho_{F,0}(0)\rangle = (\rho_F^{(0)})^2 -
  \frac{\eta^{(F)}_{ADW}}{4\pi^2 x^2},    
\end{eqnarray}
where we have defined $\rho_{F,0}(x)=(\psi^+)^\dagger(x) \psi^+(x) +
(\psi^-)^\dagger(x) \psi^-(x)$. 
For the bosonic density wave fluctuations, using
the Haldane expansion of the density, 
\begin{eqnarray}
  \label{eq:haldane-density-bosons}
  \rho_B(x)=\rho_B^{(0)} - \frac 1 \pi \partial_x \phi_B + \sum_{m\ge
    1} A_m \cos (2 m \phi_B - 2m \pi \rho_B^{(0)} x),  
\end{eqnarray}
we find that the density density correlation function of the bosons
reads:
\begin{eqnarray}
  \label{eq:boson-density-correlations}
  \langle  \rho_B(x)  \rho_B(0) \rangle =(\rho_B^{(0)})^2 
-
  \frac{\eta_{ADW}^{(B)}}{4\pi^2 x^2}
 +
  \sum_{m \ge 1} \frac{A_m^2 \cos
  (2m \pi \rho_B^{(0)} x)}{(x/\alpha)^{m^2 \eta_{ADW}^{(B)}}} . 
\end{eqnarray}
The dominant correlations are at wavevector $2\pi \rho_B^{(0)}$ and
are characterized by the exponent: 
\begin{eqnarray}
   \label{eq:exponent-adw-b}
  \eta^{(B)}_{ADW}=\frac{2}{u_++u_-}\left[M_{BB}+\frac{u_+
      u_-}{\mathrm{det}(N)} N_{FF}\right] .
\end{eqnarray}
This exponent is obtained from the fermionic exponent by the exchange
$(M_{BB},N_{FF}) \leftrightarrow (M_{FF},N_{BB})$.  The exponent is also
divergent when $u_- \to 0$.  

It is also interesting to consider the cross correlations between
bosonic and fermionic density. One has:
\begin{eqnarray}
  \label{eq:cross-density}
  \langle \rho_B(x) \rho_F(0)\rangle = \rho_B^{(0)} \rho_F^{(0)}
  -\frac{M_{BF}-u_+ u_- N_{BF}/\mathrm{det}(N)}{ 
    u_++u_-} \frac 1 {2 \pi^2 x^2},   
\end{eqnarray}
so that the non-uniform  components of the densities of bosons and
fermions remain uncorrelated. The cross density  correlations 
vanish when $V_{BF}=0$ and are positive when
$V_{BF}<0$ as a result of the boson-fermion attraction.

 For finite temperature, the
correlation
functions of Eqs.\ (\ref{eq:fermion-density-correlations})
and
(\ref{eq:boson-density-correlations})
decay exponentially, with correlation lengths given by:
\begin{eqnarray}
  \label{eq:adw-thermal-lengths}
  \xi_{\mathrm{ADW}}^{(F)}=\frac{2 \mathrm{det}(N)}{\pi N_{BB} T} ,
  \\ 
   \xi_{\mathrm{ADW}}^{(B)}=\frac{2 \mathrm{det}(N)}{\pi N_{FF} T} .
\end{eqnarray}

\subsection{superfluid correlations}
\label{sec:superfluid}

Fermions, due to their spinless character, can only present p-wave
type superfluidity. The order parameter is 
$-i \psi(x) \nabla \psi(x)$ and can be expressed using the
decomposition of Eq.\ (\ref{eq:fermion-bosonized}) in the form:
\begin{eqnarray}
  \label{eq:fermion-sf-op}
  \psi_F^+(x) \psi_F^-(x) \sim \frac{e^{2i \theta_F}}{2\pi \alpha}.   
\end{eqnarray}
The order parameter for p-wave superfluidity exhibits 
algebraic correlations:
\begin{eqnarray}
  \langle \psi_F^+(x) \psi_F^-(x)  (\psi_F^+(0) \psi_F^-(0))^\dagger
  \rangle \sim (x/\alpha)^{-\eta_{S}^{(F)}},  
\end{eqnarray}
with the  exponent for fermion superfluidity:
\begin{eqnarray}
  \label{eq:fermion-sf-exponent}
 \eta_{S}^{(F)}=\frac{2}{u_++u_-}\left[N_{FF}+\frac{u_+
     u_-}{\mathrm{det}(M)} M_{BB}\right] .
\end{eqnarray}
We now turn to the superfluid fluctuations of the bosons. 
The quasi-long range superfluid order is characterized by the
correlation function:
\begin{eqnarray}
  \label{eq:superfluid-correl}
  \langle \psi_B^\dagger(x) \psi_B(0) \rangle 
 \sim \langle {\rm e}^{- i \theta_B(x) + i \theta_B(0)} \rangle 
\sim (x/\alpha)^{-\eta^{(B)}_S}.  
\end{eqnarray}
Our result for the superfluid exponent  $\eta^{(B)}_S$ is: 
\begin{eqnarray}
  \label{eq:exponent-super}
  \eta^{(B)}_S=\frac{1}{2(u_++u_-)}
  \left[N_{BB}+\frac{u_+u_-}{\mathrm{det}(M)} M_{FF} \right],    
\end{eqnarray}
  
The superfluid and the density wave exponents are not independent from
each other. Indeed, noting that $u_+ u_-=\mathrm{det}(MN)^{1/2}$, we
have that: $4 \eta^{(B)}_S=(\mathrm{det}(N)/\mathrm{det}(M))^{1/2}
\eta_{ADW}^{(F)}$ and  $\eta^{(F)}_S=(\mathrm{det}(N)/\mathrm{det}(M))^{1/2}
\eta_{ADW}^{(B)}$. This implies that the exponents satisfy the
relation $\eta_{ADW}^{(F)} \eta_{S}^{(F)} = 4 \eta_{ADW}^{(B)}
\eta_{S}^{(B)}$.

Turning to the finite temperature case, the thermal lengths are: 
\begin{eqnarray}
  \label{eq:sf-thermal-lengths}
  \xi_{\mathrm{S}}^{(F)}=\frac{2 \mathrm{det}(M)}{\pi M_{BB} T} ,
  \\ 
   \xi_{\mathrm{S}}^{(B)}=\frac{2 \mathrm{det}(M)}{\pi M_{FF} T} .
\end{eqnarray}

\section{Bragg scattering} \label{sec:bragg-scattering}
According to Ref.\ \onlinecite{brunello_bragg}, the Bragg scattering intensity is
proportional to the imaginary part of the retarded density-density response
functions $\chi(q,\omega)$.   Retarded density response functions
$\chi_{ab}(q,\omega)$ ($a,b=B,F$) can be obtained from the Fourier
transform of Green's
functions~(\ref{eq:Green1}) as: 
\begin{eqnarray}
  \label{eq:density-response}
  \chi_{ab}(q,\omega) =-\frac{q^2}{\pi^2}
  G_{ab}(q,\omega_n)|_{i\omega_n \to \omega+ i0_+}.  
\end{eqnarray}   
 Using the expression of the Fourier transform
 (\ref{eq:green-phi-Fourier}) from Sec.~\ref{sec:correlations}
  we obtain for $\omega>0$ : 
 \begin{eqnarray}
   \label{eq:im-chi}
    \mathrm{Im}  \chi_{ab}(q,\omega) &=& - \frac{|q|}{2} \left[
      \frac{u_-(M_{ab}-u_+^2(N^{-1})_{ab})}{u_+^2-u_-^2}
      \delta(\omega-u_-|q|) \right.\nonumber \\ &+&\left.  \frac{u_+(u_-^2(N^{-1})_{ab}-M_{ab})}{u_+^2-u_-^2}
     \delta(\omega-u_+|q|) \right].
 \end{eqnarray}
Equation (\ref{eq:im-chi}) predicts peaks at
frequencies $u_\pm |q|$. The matrices $M$ and $N$ can be deduced from
the spectral weight of these peaks. One has the following sum rules:
\begin{eqnarray}
  \int_0^\infty d\omega\mathrm{Im} \chi_{ab}(q,\omega) &=& \frac{|q|}{2\pi}
  (\eta_{\phi})_{ab}, \\
  \int_0^\infty \frac{d\omega}{\omega}  \mathrm{Im}
  \chi_{ab}(q,\omega) &=& \frac 1 2 (N^{-1})_{ab}, \\ 
     \int_0^\infty d\omega \omega  \mathrm{Im}
  \chi_{ab}(q,\omega) &=& \frac{q^2} 2 M_{ab}.     
\end{eqnarray}
The first sum rule is simply a restatement of our derivation of the
equal time Green's function in Sec.~\ref{sec:hamiltonian}. 
The second sum rule is a Kramers-Kronig
relation giving the real part of the zero frequency 
(matrix) density-density response  function as an integral 
of its imaginary part. Since the real part of the density density
response function is the compressibility, the second result is not
surprising. The last sum rule is a consequence of current
conservation. Indeed, using current conservation, one can relate the
density-density response function to the current-current response
function. Using again a Kramers-Kronig, the last integral is shown to
be equal to the static current-current response function.   

{
\section{Variation of exponents} \label{sec:toy-model}
}  

In this section, we 
first 
consider the model of
Eq.~(\ref{eq:bosonized-perturb}) with $M_{BF}=0$. 
 For the two-component case ($a,b \in \{B,F\}$), 
the velocities are found as \cite{cazalilla03_mixture}: 
 \begin{eqnarray}
   \label{eq:velocities}
   u_\pm^2 = \frac{u_F^2+u_B^2} 2 \pm \sqrt{\left( \frac{u_F^2-u_B^2}
       2\right)^2 
+  \left(\frac{V_{BF}}\pi \right)^2 u_B K_B u_F
     K_F
}.
 \end{eqnarray}
Using the Green's function methods of Sec.~\ref{sec:hamiltonian}, we
obtain the following expressions for the exponents: 
\begin{eqnarray}\label{eq:adw-b-toy} 
  \eta_{ADW}^{(B)}&=&\frac{u_B K_B}{u_+ u_-} \left[ u_+ + u_- +
    \frac{u_F^2-u_B^2}{u_++u_-}\right],  \\
    \label{eq:adw-f-toy} 
   \eta_{ADW}^{(F)}&=&\frac{u_F K_F}{u_+ u_-} \left[ u_+ + u_- - 
    \frac{u_F^2-u_B^2}{u_++u_-}\right], \\
    \label{eq:sf-b-toy} 
   \eta_{S}^{(B)}&=&\frac 1 {4 u_B K_B} \left[ u_+ + u_- -
    \frac{u_F^2-u_B^2}{u_++u_-}\right], \\
     \label{eq:sf-f-toy} 
   \eta_{S}^{(F)}&=&\frac 1 {u_F K_F} \left[ u_+ + u_- + 
    \frac{u_F^2-u_B^2}{u_++u_-}\right]. 
\end{eqnarray}
Finally, using that $u_+^2+u_-^2=u_F^2+u_B^2$ and $u_+^2 u_-^2=u_F^2
u_B^2 - (V_{BF} u)^2/\pi^2$
where $u\equiv \sqrt{u_FK_F u_B K_B}$,
 we can express the exponents entirely as
functions of $u_F,u_B,K_F,K_B$ and $V_{BF}$. 
Expanding to second order in $V_{BF}$ we find:
\begin{eqnarray}
  \eta_S^{(B)}&=&\frac 1 {2K_B} -\frac{V_{BF}^2 K_F u_F}{4\pi^2 u_B
    (u_F+u_B)^2} + O(V_{BF}^4), \\
 \eta_S^{(F)}&=&\frac 2 {K_F} -\frac{V_{BF}^2 K_B u_B}{\pi^2 u_F
    (u_F+u_B)^2} + O(V_{BF}^4), \\
\eta_{ADW}^{(B)}&=&2K_B\left[1 + \frac{V_{BF}^2 K_F
    K_B(2u_B+u_F)}{2\pi^2(u_F+u_B)^2 u_B}+ O(V_{BF}^4) \right],
\nonumber \\ \\  
\eta_{ADW}^{(F)}&=& 2K_F\left[1 + \frac{V_{BF}^2 K_F
    K_B(2u_F+u_B)}{2\pi^2(u_F+u_B)^2 u_F}+ O(V_{BF}^4)
\right].\nonumber \\    
\end{eqnarray}
This shows that superfluidity is enhanced by the boson-fermion
interaction. The enhancement of fermionic superfluidity  
becomes weaker as $u_F$ is increased. By contrast,  when $u_F$ increased,
the enhancement of bosonic  superfluidity is non monotonous, being
maximum for $u_F=u_B$. At the same time, the exponent of density wave
correlations for the bosons and fermions are increased by $V_{BF}$, 
indicating a reduction of the density wave quasi-long range order.

We note that the density-wave exponents
Eqs.~(\ref{eq:adw-b-toy})--(\ref{eq:adw-f-toy})  can be
rewritten respectively as  $ \eta^{(B)}_S=(u_B^2+u_+
u_-)/(2u_B K_B \sqrt{u_B^2+u_F^2+2u_+u_-})$ and  $\eta^{(F)}_S=(u_F^2+u_+
u_-)/[u_F K_F\sqrt{u_B^2+u_F^2+2u_+u_-}]$, which are both increasing 
functions  of $u_+ u_-$. Since from Eq.\ (\ref{eq:velocities}), $u_- u_+$ 
is a decreasing function of  $|V_{BF}|$, the superfluid exponents are  
decreasing functions of $|V_{BF}|$. In other words, boson-fermion
interactions lead to an enhancement of  superfluid
fluctuations.  

In the context of electrons coupled
to acoustic phonons, an analogous enhancement of fermionic superfluidity was
already noticed \cite{martin1995}. The interpretation of the
result~(\ref{eq:fermion-sf-exponent}) is that the sound modes of the
bosonic fluid interact with the fermionic atoms exactly as the  acoustic
phonons interact with the electrons in the model of
Ref.~\onlinecite{martin1995}. The interaction of the acoustic modes
with the fermions  gives rise to an effective attractive
interaction  that  enhances superfluidity. The enhancement of bosonic
superfluidity can also be understood in terms of effective
attraction between bosons generated by the sound modes of the Fermi
fluid.  We note that in the
context of electronic systems, the collapse or phase separation that
results from excessively strong $V_{BF}$ interaction is known as the
Wentzel-Bardeen instability \cite{wentzel1951,bardeen1951}. The
enhancement of superfluid correlations in mixtures 
could be tested by varying the
strength of the boson-fermion interaction $V_{BF}$ in a
${}^{40}$K--${}^{87}$Rb mixture using a Feshbach
resonance \cite{modugno2007}. Superfluid correlations will present a
minimum when $V_{BF}$ is tuned to zero by interaction.  

\begin{figure}
  \centering
\includegraphics[width=8cm]{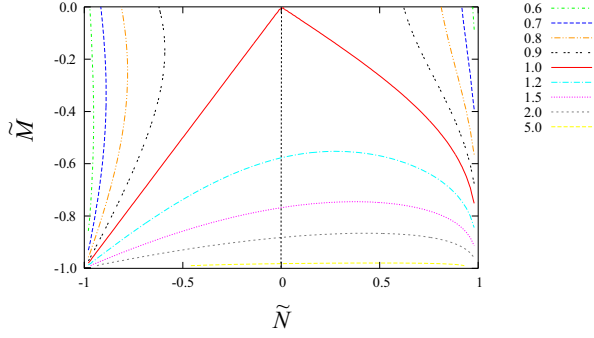}   
  \caption{
The contour plot of  $2K_B\eta_S^{(B)}$ with the fixed  $\xi=u_F/u_B=2$ 
on the plane of 
{
 $\tilde{N}$ and  $\tilde{M}(< 0)$ which are the 
 dimensionless quantities defined by 
$\tilde{N} \equiv N_{BF} \sqrt{K_BK_F}/\sqrt{u_Bu_F}$ and 
$ \tilde{M} \equiv M_{BF}(\sqrt{K_BK_F})^{-1}/\sqrt{u_Bu_F}$, respectively. 
}
{For $\tilde{N}=0$ ($\tilde{M} = 0$),
 the quantity  $2K_B\eta_S^{(B)}$ increases (decreases) monotonously  as 
 $|\tilde{M}|$ ($|\tilde{N}|$) increases.} 
}
\label{fig:fig3}
\end{figure}

Next,   we discuss
  the case in the presence of both 
 $N_{BF} (= V_{BF}/\pi)$ and $M_{BF}$ 
in Eq.\ (\ref{eq:bosonized}),
which give the Hamiltonian,     
\begin{eqnarray}
  \label{eq:bosonized_W}
  H&=&\sum_{\nu=F,B} \int \frac{dx}{2\pi} \left[ u_\nu K_\nu(\pi \Pi_\nu)^2 +
    \frac{u_\nu}{K_\nu} (\partial_x
    \phi_\nu)^2\right] \\ &+& \frac{N_{BF}}{\pi}
 \int dx  \partial_x
  \phi_B \partial_x \phi_F +
  \pi M_{BF} \int dx  \Pi_B \Pi_F.  \nonumber  
\end{eqnarray}
As shown in Appendix \ref{app:galilean}, 
 we have the constraint $M_{BF}<0$ from the {Galilean invariance} of
  the system (see Eq.\ (\ref{eq:m-response-bf})),  so 
we restrict ourselves to  the case of $M_{BF}<0$
 in the following.
The velocity  is obtained as 
\begin{widetext}
\begin{eqnarray}
  \label{eq:velocities_w}
  u_\pm^2 = \frac{u_F^2+u_B^2}{2} 
    +   N_{BF} M_{BF}   
\pm \sqrt{\left( \frac{u_F^2-u_B^2}{2}\right)^2 
 +      u_B K_B u_F K_F
 \left(N_{BF} + \frac{M_{BF}u_{B}}{K_{B}K_{F}u_{F}} \right) 
  \left(N_{BF} + 
\frac{M_{BF}u_{F}}{K_{B}K_{F}u_{B}}\right)   
} .
\end{eqnarray}
\end{widetext}
Thus the excitation is stable for 
\begin{eqnarray}
   \label{eq:u_N_M}
 u_+^2 u_-^2 = (u_Bu_F - M_{BF}^2 /(K_BK_F)) (u_Bu_F - N_{BF}^2K_BK_F) > 0 .
\end{eqnarray}
From Eq.\ (\ref{eq:exponent-super}),
the exponent for the superfluidity for boson is calculated as
\begin{widetext} 
\begin{eqnarray}
  \eta^{(B)}_{S}&=&
\frac{ 1/(2u_B K_B)}
{\sqrt{u_B^2 + u_F^2 +2 N_{BF}M_{BF} + 2u_+u_-}}
 \left[
  u_B^2 +\frac{u_Bu_F(u_Bu_F - N_{BF}^2K_BK_F)}
{u_+ u_-}
\right] 
\nonumber \\
&=&
\frac{1}{2K_B}
\frac{ 1 }
{\sqrt{1 + \xi^2 +2 \xi \tilde{N} \tilde{M}
  + 2\xi \sqrt{(1-\tilde{N}^2)(1-\tilde{M}^2)}}}
 \left(
  1 + \xi \sqrt{\frac{1 - \tilde{N}^2}{1-\tilde{M}^2} }
\right) .
   \label{eq:exponent-adw-b_w}
\end{eqnarray}
\end{widetext}
where
$\tilde{N}=N_{BF} \sqrt{K_BK_F}/\sqrt{u_Bu_F}$
{ and }
$\tilde{M}=M_{BF}(\sqrt{K_BK_F})^{-1}/\sqrt{u_Bu_F}$. 
 The second line shows that  $2K_B\eta_S^{(B)}$ depends on 
 $\tilde{N}$,  $\tilde{M}$, and $\xi = u_F/u_B$. 
The stable conditions for $u_{\pm}$ are expressed as
 $-1<\tilde{N},\tilde{M}<1$.
 At the boundary of $|\tilde{M}| = 1$, $\eta^{(B)}_{S}$ becomes infinite  
 while it stays constant at $|\tilde{N}| = 1$ $(|\tilde{M}| \not= 1)$.
Since  $\tilde{N}$ ($\tilde{M}$)
    has an effect of decreasing  (increasing)  $\eta^{(B)}_{S}$, 
they compete with each other.
 In Fig.~\ref{fig:fig3}, the contour plot of 
the quantity $2K_B\eta_s^{(B)}$ is shown 
on the plane of $\tilde{N}$ (horizontal axis) and 
$\tilde{M}$ (vertical axis) with the fixed 
$\xi=2$,
 where the case of $2K_B\eta_s^{(B)}<1 (>1)$ denotes the enhancement (suppression)
 of the superfluidity.
{
 For $\tilde{N}=0$, 
 the current-current boson-fermion coupling  $\tilde{M}$ suppresses 
  the superfluidity.
 The effect of the suppression reduces and vanishes, i.e.,  
  $\eta_s^{(B)}=1/(2K_B)$
on the solid line 
in Fig.\ \ref{fig:fig3}
}
  due to the compensation effect of  these two coupling constants. 
{
 For  small  $|\tilde{M}|$ and $|\tilde{N}|$,    
  $ \eta_s^{(B)}=1/(2K_B)$ is obtained  
  at $\tilde{M}=\tilde{N}$ for  $\tilde{M} \tilde{N} > 0$, and  
  at $\tilde{M} = - \tilde{N} \xi/(1+\xi) $ for $\tilde{M} \tilde{N} < 0$.
}
In the region of 
$\tilde{N}<0$,
we find a novel feature  for small  $|\tilde{M}|$. 
The superfluidity is enhanced  by the coupling $|\tilde{N}|$,
and is further enhanced  due to the coupling $|\tilde{M}|$. 
The enhancement becomes optimized, i.e.,
  the minimum value of $2K_B \eta_s^{(B)}$ is obtained
   at a certain value of $|\tilde{M}|$. 


\section{Conclusion} 
We have analyzed the mixture of bosonic and fermionic atoms in one
dimension using a Green's function equation of motion method starting
from a phenomenological bosonized Hamiltonian. We have derived
expressions of the zero temperature exponents and of the thermal
correlation thermal correlation lengths for the density wave and
superfluid order parameters.   
For the exponents, we have 
 recovered the exponents previously obtained for the mixture
in Ref.\ \onlinecite{mathey2007a} and we have studied their behavior as a
function of the fermion density for fixed interaction. W
e have found that for weak
interaction, the behavior of the superfluid exponent of the bosons
could be non-monotonous, although boson superfluid correlations were
always enhanced with respect to the system without fermions.  
Such behavior can lead to a non-monotonous
dependence of the superfluid transition temperature as a function of
fermion density in an array of
weakly coupled mixtures generalizing the bosonic array considered in
Ref.\ \onlinecite{cazalilla_deconfinement_longpaper}. 
 This
behavior is also in contrast with the one observed in experiments with three
dimensional interacting systems \cite{guenter2006}, where fermion
doping was seen to reduce superfluidity. In one dimensional systems at
incommensurate filling, fermion doping is seen to always enhance
superfluidity, the maximal enhancement being obtained near the
collapse instability.
By contrast, we have seen 
that density wave exponents are decreasing functions of the fermion
density, which recover their  value in the absence of fermion-boson
interaction only in the limit of large fermion density.  We have also been able to
predict how the relative weights of the peaks in Bragg scattering 
intensity depend on the parameters of the phenomenological
Hamiltonian.

\begin{acknowledgments}
E. O. acknowledges support and hospitality from the Graduate School of
Science and the Physics Department of Nagoya University in June 2009, 
when part of this work was completed.   
\end{acknowledgments}

\appendix

\section{Consequences of Galilean invariance} 
\label{app:galilean}
Let us consider a system with {$N_p$} different types of particles indexed
by $1\le a\le {N_p}$ with masses {$M_a$.} 
Under a Galilean
boost,  $p_{i,a} \to p_{i,a} + {M_a} v$ and $r_{i,a} \to r_{i,a} + vt$. 
This transformation is realized by the unitary operator:
\begin{eqnarray}
  \label{eq:galilean-boost}
  U=\exp\left[i \sum_{a=1}^{{N_p}} \sum_{i_a=1}^{N_a} M_a v r_{i,a}\right]  
\end{eqnarray}
One has: 
\begin{eqnarray}
  \label{eq:galilean-ham}
  U^\dagger H U = H + v P + \frac 1 2 M v^2, 
\end{eqnarray}
where {the total mass} $M=\sum_a N_a M_a$ and {the total momentum}:
\begin{eqnarray}
  P=\sum_{a=1}^{{N_p}} \sum_{i_a=1}^{N_a} p_{i_a}, 
\end{eqnarray}
is the total momentum of the system.
In second quantization, the operator $U$ takes the form:
\begin{eqnarray}
  U=\exp\left[i \int dr \sum_{a=1}^{{N_p}} {M_a} v r \rho_a(r)  \right],  
\end{eqnarray}
and one has : $U^\dagger \psi_a(r) U = e^{i {M_a} v r}
\psi_a(r)$. Therefore, in bosonization, a Galilean boost takes the
form $\theta_a(r) \to \theta_a(r) + {M_a} v r$, i.e., $\pi \Pi_a(r) \to
\pi \Pi_a(r)  + {M_a} v$.   
 From the conservation
equation for particles of type $a$, $\partial_x j_a + \partial_t
\rho_a=0$ and the bosonization relations~(\ref{eq:haldane-density-bosons}), we
have that the current density is $j_a=\partial_t \phi_a/\pi$. Using
the equation of motion, we can rewrite $j_a=\sum_b M_{ab} \nabla
\theta_b/\pi$. Therefore, under a Galilean boost, we will have:
\begin{eqnarray}
  \langle j_a \rangle = \frac 1 \pi \sum_b M_{ab} {M_b} v. 
\end{eqnarray}
Turning to the particle current, in the rest frame, the particle
current will be $\langle j_a \rangle= v \rho_a$. 
This can be seen for instance by
calculating $j_a=\partial_t (\phi_a(x-vt))/\pi$. From this result, 
it is clear that one must have:
\begin{eqnarray}
  \pi \rho_a = \sum_b  M_{ab} {M_b}, 
\end{eqnarray}
for each $a$. In the case $N_p=2$ this gives
Eq.~(\ref{eq:galilean-constraint}). 

An alternative derivation of Eq.\ (\ref{eq:galilean-constraint}) 
 is obtained by calculating the ground state
energy under twisted boundary conditions~(\ref{eq:definition-M}) 
 using second order perturbation theory \cite{kohn_stiffness}.
One finds:
 \begin{eqnarray}
   \delta E = \sum_a \frac{\rho_a}{2 {M_a} L} \phi_a^2 + \frac 1 {L^2} 
   \sum_{a,b,n} \frac{\langle 0|J_a|n \rangle \langle n | J_b
     |0\rangle}{E_n -E_0} \varphi_a \varphi_b,    
 \end{eqnarray}
where $J_a=P_a/{M_a}$ is the current of particles of type $a$.  
Because of the translational invariance of the system, the eigenstates
of the Hamiltonian are also eigenstates of the 
total momentum operator $P=\sum_a P_a$.  Since the ground state is
non-degenerate, this implies that $\langle n | P |0\rangle=0$, i.e., for the 
Bose-Fermi mixture: $\langle 0|P_F|n \rangle +  \langle 0|P_B|n
\rangle=0$, leading to the  following expressions for
$M_{ab}$: 
{
\begin{eqnarray}\label{eq:m-response} 
  M_{BB}&=&\frac{\pi \rho_B}{M_B} + \frac{2\pi}{L}
  \frac{\chi}{M_B^2}, \\
  M_{FF}&=& \frac{\pi \rho_F}{M_F} + \frac{2\pi}{L}
  \frac{\chi}{M_F^2}, \\ 
 M_{BF}&=& - \frac{2\pi}{L}
  \frac{\chi}{M_F M_B},  
\label{eq:m-response-bf} 
\end{eqnarray} } 
where:
\begin{eqnarray}
  \chi=\sum_n \frac{|\langle n |P_B|0\rangle|^2}{E_n -E_0}.  
\end{eqnarray}
Obviously, $\chi>0$. Using the expressions (\ref{eq:m-response}),   
we obtain: 
\begin{eqnarray}
{\mathrm{det}(M)=\frac{\pi^2 \rho_F \rho_B}{M_F M_B} + \left(\frac{\pi \rho_F
}{M_F M_B^2} + \frac{\pi \rho_B
}{M_B M_F^2}\right) \chi,}       
\end{eqnarray}
so that $\mathrm{det}(M)>0$. 
 

\end{document}